\documentclass{article}
\usepackage{amsmath,graphicx}
\usepackage[preprint]{spconf}

\usepackage{cite} 
\usepackage{xcolor}
\usepackage{amsmath,amssymb,amsfonts}
\usepackage{booktabs}
\usepackage{multicol, multirow}
\usepackage{makecell}
\usepackage{array}
\usepackage{bbm}
\usepackage{empheq}

\copyrightnotice{\copyright\ IEEE}
\toappear{To appear in {\it Proc.\ of IEEE ICASSP, 2023}}

\title{AD-YOLO: You Look Only Once in training \\ Multiple Sound Event Localization and Detection}
\name{Jin Sob Kim, Hyun Joon Park, Wooseok Shin, and Sung Won Han$^\ast$
\thanks{This research was supported by Brain Korea 21 FOUR. This research was also supported by Korea University Grant (K2107521,K2202151) and a Korea TechnoComplex Foundation Grant (R2112651).}
}
\address{School of Industrial and Management Engineering, Korea University, Seoul, Republic of Korea}

\begin{document}
%
\maketitle
\begin{abstract}
Sound event localization and detection (SELD) combines the identification of sound events with the corresponding directions of arrival (DOA). 
Recently, event-oriented track output formats have been adopted to solve this problem; however, they still have limited generalization toward real-world problems in an unknown polyphony environment.
To address the issue, we proposed an angular-distance-based multiple SELD (AD-YOLO), which is an adaptation of the ``You Only Look Once" algorithm for SELD.
The AD-YOLO format allows the model to learn sound occurrences location-sensitively by assigning class responsibility to DOA predictions.
Hence, the format enables the model to handle the polyphony problem, regardless of the number of sound overlaps.
We evaluated AD-YOLO on DCASE 2020-2022 challenge Task 3 datasets using four SELD objective metrics. 
The experimental results show that AD-YOLO achieved outstanding performance overall and also accomplished robustness in class-homogeneous polyphony environments.
\end{abstract}

\section{Introduction}
\label{sec:introduction}

Given a multi-channel audio input, sound event localization and detection (SELD) combines sound event detection (SED) along the temporal progression and the identification of the direction-of-arrival (DOA) of the corresponding sounds.
The combining of two sub-tasks can perform a key role in automatic audio surveillance \cite{crocco2016audio, chang2018feature}, navigation \cite{foggia2015audio}, and crime-safety applications \cite{valenzise2007scream}.
Recently, neural-network-based approaches to solving the SELD problem have made progress via several competitions such as the DCASE challenge \cite{politis2020overview}.
As many prior studies \cite{adavanne2018sound, cao2019polyphonic, mazzon2019first, yasuda2022echo, cao2021improved, shimada2021accdoa, shimada2022multi} have been conducted on SELD, various approaches have been proposed to tackle the problem.

Several works \cite{adavanne2018sound, cao2019polyphonic, mazzon2019first, yasuda2022echo, cao2021improved} have adopted a two-branch output format, considering SELD as the performing of two separate sub-tasks from each branch, SED and DOA (SED-DOA).
On the other hand, \cite{shimada2021accdoa, shimada2022multi} solve the task in a single-branch output through a Cartesian unit vector (proposed as ACCDOA \cite{shimada2021accdoa}), combining SED and DOA representations, where the zero-vector represents none. 
However, all have in common adopting a track output format which is an event-oriented approach.

In terms of handling the polyphony, \cite{adavanne2018sound, cao2019polyphonic, yasuda2022echo, shimada2021accdoa} allow a single track per event class. Therefore, it is clear that the class-wise output format has limitations when same-class overlapping occurs. 
In the meantime, \cite{mazzon2019first, cao2021improved} design the network architectures to output the desired number of tracks regardless of event class, where one and two are set for the maximum overlaps respectively.
\cite{shimada2022multi} suggests allowing multiple tracks, implementally up to three, per event class and utilizing class-wise track permutation learning (ADPIT).

Given the labeled data, the maximum polyphony can be considered, however, the maximum is priorly unknown under the real-world circumstance or the unlabeled.
Thus, the event-oriented track output formats intrinsically contain the limitation of presetting the number of tracks, constraining the generality and expandability of the method itself.

In this paper, we propose a novel approach for SELD, undisturbed by the unknown number of polyphony by performing location-oriented detection, out of the event-oriented perspective. 
We adapt the framework of "You Only Look Once" (YOLO) \cite{Redmon_2016_CVPR}, renowned for multiple object detection from images, to the SELD by using the notion of angular distance, namely proposing angular-distance-based YOLO (AD-YOLO).
The results of an experiment using the series of DCASE 2020-2022 Task 3 (SELD) datasets \cite{politis2020dataset, politis2021dataset, politis2022starss22} demonstrated that AD-YOLO outperformed existing SELD formats in both overall evaluation and polyphonic circumstances.

\section{Proposed Method}
\label{sec:method}

\subsection{Training SELD in YOLO}
\label{subsec:training}

\begin{figure}[!t]
\centerline{\includegraphics[scale = 0.48]{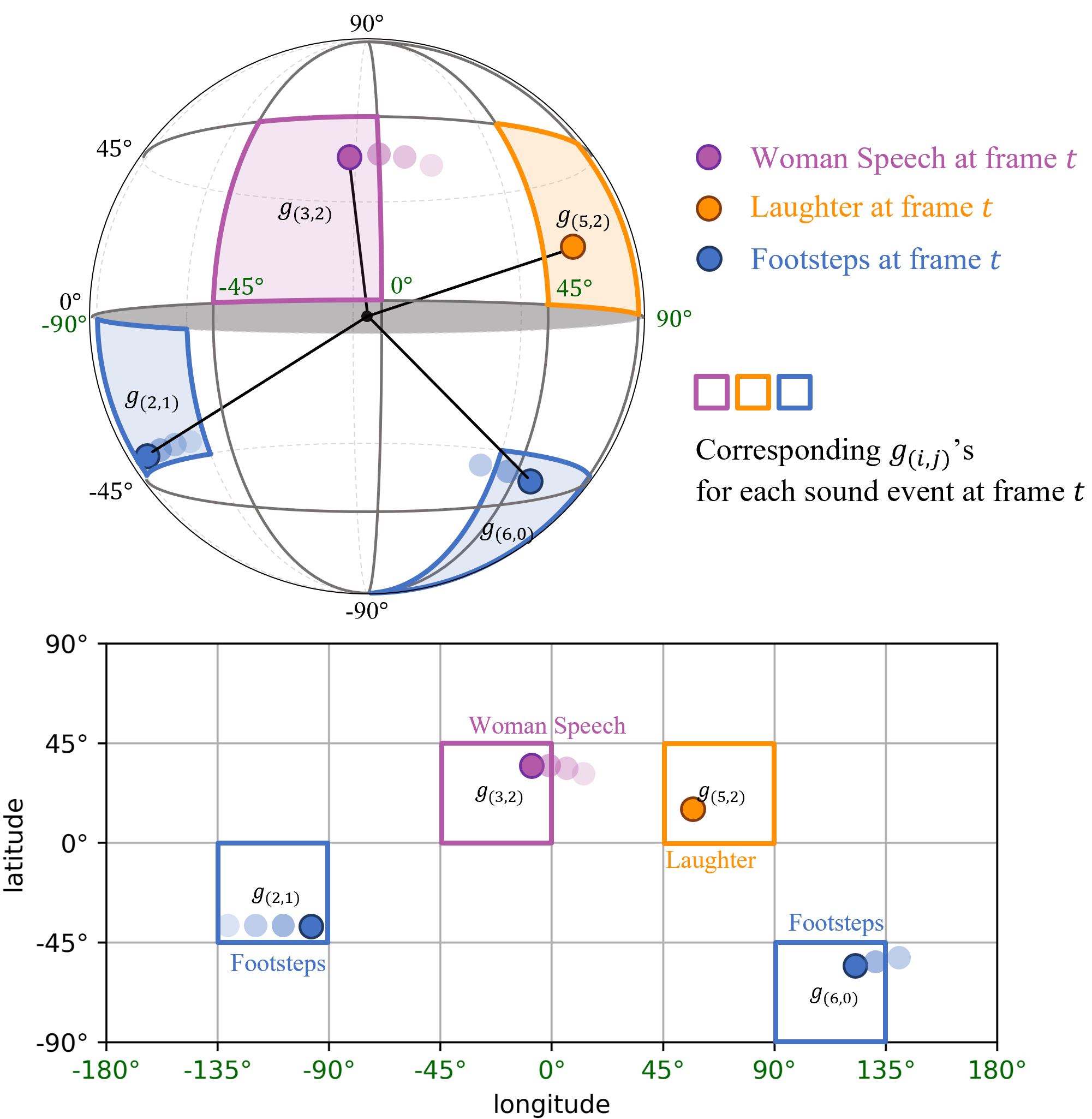}}
\caption{Along the order of transparency, the figure illustrates the trajectory of DOA from time $t-3$ to $t$ with the identification of class activity. Given $(45^{\circ}, 45^{\circ})$ of unit space, each grid corresponding to the sound events at the current time $t$ can be mapped as shown above.}
\label{fig:SELD-YOLO-format}
\end{figure}

Given a sound source corresponding to $\mathbf{T}$ time frames, we let $M$ sound event targets exist at the frame $t$.
The $m^{th}$ reference target, among $M$ events at any time $t$, can be defined by $\{ c_{m}, \lambda_{m}, \phi_{m} \}$, where $c_{m}$ denotes the class of the corresponding sound, and ($\lambda_{m}, \phi_{m}$) tuple, represented by a polar coordinate system, indicate the location of the sound source at time $t$. By dividing a spherical surface into grids, any polar system coordinate can have a corresponding area; therefore, we redefine the format using the indices of the grid. 
$\{g_{(i,j)}, c_{m}, \lambda_{m}, \phi_{m} | \lambda_{m} \rightarrow i, \phi_{m} \rightarrow j \}$, where $g_{(i,j)}$ indicates the area depending on $\lambda_{m}$ and $\phi_{m}$ among the spatially-uniformly divided areas in terms of longitude and latitude.
Then, we can depict the reference data as $\mathbf{R} \in \mathbb{R}^{\mathbf{M} \times 5}$, where $\mathbf{M}=\sum_{\forall t} M$, and each row contains the indicator of time $t$, corresponding grid $g_{(i,j)}$, active class $c$, and DOA $(\lambda, \phi)$. 
Fig. \ref{fig:SELD-YOLO-format} shows an example of the reference label tagged with the corresponding grid areas on the spherical surface.

We let the backbone network $\mathcal{F}(\mathbf{X}; \theta)$ transform the multi-channel acoustic feature $\mathbf{X} \in \mathbb{R}^{Q \times T' \times f}$ into the embedding space representation $x \in \mathbb{R}^{\mathbf{T} \times d}$ with the parameters $\theta$.
$Q$, $T'$, and $f$ denote the number of channels, time frames, and dimensions of the acoustic feature input $\mathbf{X}$ respectively, and $d$ is the size of the embedding space.
Then, $x$ is followed by a single fully connected layer (FC), estimating both SED and DOA with the parameters $\theta_{\mathrm{FC}}$.
\begin{equation}
\begin{aligned}
\begin{split}
x &= \mathcal{F}(\mathbf{X}; \theta) \\
\hat{y} &= \mathrm{FC}(x; \theta_{FC})
\end{split}
\end{aligned}
\label{eq:model-processing}
\end{equation}
where the estimation $\hat{y} \in \mathbb{R}^{\mathbf{T} \times [\mathbf{G} \times \mathbf{K} \times (\mathbf{C}+3)]}$ consists of $\mathbf{K}$ predictions for $\mathbf{G}$ grids, and each prediction contains $\mathbf{C}$ class-wise confidence scores, the score for sound existence, and the DOA prediction $(\hat{\lambda}, \hat{\phi})$.

\begin{figure}[!t]
\centerline{\includegraphics[scale = 0.48]{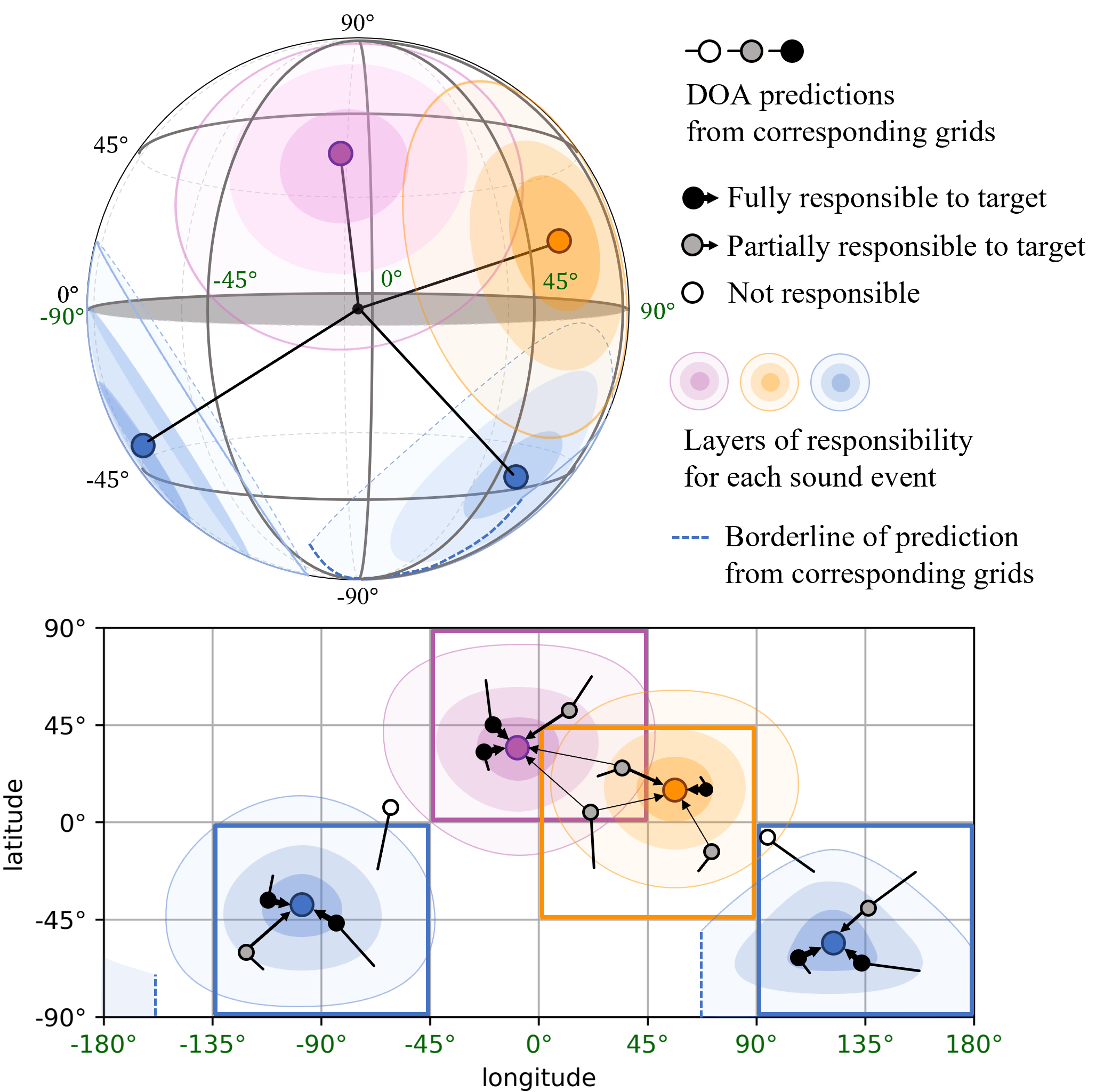}}
\caption{Finding responsible during training at frame $t$ in Fig \ref{fig:SELD-YOLO-format}. Given a set of $\tau = \{ 45^{\circ}, 25^{\circ}, 10^{\circ} \}$, multi-layered responsible boundaries are shown. Different levels of responsibilities are assigned to each prediction from corresponding grids.}
\label{fig:RESPONSIBLES}
\end{figure}
During training, we determine the responsible predictions for each reference in $\mathbf{R}$ using an angular distance $\delta$ as follows:
\begin{equation}
\small
\begin{aligned}
\delta_{mk} = \mathrm{cos}^{-1}\big[&\mathrm{sin}(\hat{\phi}_{mk})\mathrm{sin}(\phi_{m}) \\
                                    &+\mathrm{cos}(\hat{\phi}_{mk})\mathrm{cos}(\phi_{m})\mathrm{cos}(\hat{\lambda}_{mk}-\lambda_{m}) \big]
\end{aligned}
\label{eq:angular-distance-metric}
\end{equation}
where $m$ indexes the rows from the matrix $\mathbf{R}$, $(\lambda_{m}, \phi_{m})$ corresponds to referenced DOA, and $(\hat{\lambda}_{mk}, \hat{\phi}_{mk})$ is the estimated DOA of $k^{\mathrm{th}}$ prediction from the reference time $t_{m}$ and grid $g_{m}$.
Given a training threshold $\tau$, responsibility is assigned to the prediction if $\delta_{mk} < \tau$.
We use multiple thresholds to layer the responsibility according to the distance.
Also, the overlaps between grid spaces are allowed to track the continuity of DOA trajectories crossing multiple areas.
Fig. \ref{fig:RESPONSIBLES} shows an example of responsibility assignment, where $\mathbf{K}=1$ and the allowance for overlapping grid was set to 50\%. 
As shown in the figure, the allowance extends the corresponding grids of each target, and the thickness of the arrow from the predictions indicates the level of responsibility toward each target. 

The loss function for training the network can be formulated based on the identification of responsibilities.
First, we compose the logical matrix $\mathbbm{1}_{m;\tau} \in \mathbb{R}^{\mathbf{T} \times \mathbf{G} \times \mathbf{K}}$ as follows:
\begin{equation}
\small
\begin{aligned}
    \mathbbm{1}^{i}_{m;\tau} = \left\{ 
    \begin{aligned}
        &1, \text{ if } \delta_{mi} < \tau \text{ and } (t_{m}, g_{m}) \in i \indent \\ 
        &0, \text{ otherwise} 
    \end{aligned} \right.
\end{aligned}
\end{equation}
where $i$ combines the indices $(t, g, k) \in (\mathbf{T}, \mathbf{G}, \mathbf{K})$ respectively, and $m$ indexes the $m^{\mathrm{th}}$ row from the reference matrix $\mathbf{R}$ as mentioned earlier. 

Then, the loss between responsible DOA estimations and each reference DOA is defined as follows:
\begin{equation}
\small
\begin{aligned}
l_{\delta;\tau} = \frac{\sum_{m} \sum_{i} \delta_{mi}}{\pi \sum_{m} |\mathbbm{1}_{m;\tau}|}
\end{aligned}
\end{equation}
where $|\cdot|$ denotes the number of non-zero values in the matrix.
Considering the distortion between representations on a 2-dimensional and a spherical surface, as shown in the responsible boundaries, Fig. 2, we minimize the angular distance directly rather than using the squared error of the conventional YOLO mechanism.

Subsequently, merging $\lor \mathbbm{1}_{\forall m;\tau} \in \mathbb{R}^{\mathbf{T} \times \mathbf{G} \times \mathbf{K}}$ becomes the target for the probabilistic estimation of sound source existence conditioned on its DOA, regardless of classes.
Hence, the losses can be written as:
\begin{equation}
\small
\begin{aligned}
l_{1;\tau} &= \frac{\sum_{i} \mathbbm{1}^{i}_{\forall m; \tau} \mathrm{BCE}( 1, \hat{p}_{i}(o))}{|\mathbbm{1}_{\forall m;\tau}|} \\
l_{0;\tau} &= \frac{\sum_{i} \neg \mathbbm{1}^{i}_{\forall m; \tau} \mathrm{BCE}(0, \hat{p}_{i}(o))}{|\neg \mathbbm{1}_{\forall m;\tau}|}
\end{aligned}
\end{equation}
We use binary cross entropy (BCE) for the probabilistic comparison, where $\hat{p}_{i}(o)$ denotes the estimated probability $\hat{p}(\cdot)$ of sound existence $o$ indexed by $i$ from the network output $\hat{y}$.

By further expanding the concept to class-wise responsibility, we define the logical matrix as $\mathbbm{1}_{m;\tau} \in \mathbb{R}^{\mathbf{T} \times \mathbf{G} \times \mathbf{K} \times \mathbf{C}}$ and assign 1 to $\mathbbm{1}^{i,c}_{m;\tau}$ where $c \in \mathbf{C}$ can be referenced from $c_{m} \in \mathbf{R}_{m}$.
Therefore, the classification loss for the responsible predictions is formulated as:
\begin{equation}
\small
\begin{aligned}
l_{C;\tau} = \frac{\sum_{i} \sum_{c} \mathrm{BCE}(\mathbbm{1}^{i,c}_{\forall m;\tau}, \hat{p}_{i}(c))}{\mathbf{C} |\lor_{c} \mathbbm{1}_{\forall m;\tau}|}
\end{aligned}
\end{equation}
The classification loss is averaged by the number of unique responsible predictions $\lor_{c} \mathbbm{1}_{\forall m;\tau} \in \mathbb{R}^{\mathbf{T} \times \mathbf{G} \times \mathbf{K}}$, and the merging of class-wise logical elements is denoted as $\lor_{c}$.

Finally, the formulation of the integrated loss function of AD-YOLO, layering responsibilities with a set of multiple thresholds $\tau \in \mathcal{T}$, is described as follows:
\begin{equation}
\small
\begin{aligned}
\mathcal{L}^{\text{AD-YOLO}} = \omega_{\delta}l_{\delta;\mathrm{max}(\mathcal{T})} + \frac{1}{|\mathcal{T}|}\sum_{\tau \in \mathcal{T}}
\bigg[ \omega_{1}l_{1;\tau} + \omega_{0}l_{0;\tau} + \omega_{C}l_{c;\tau} \bigg]
\end{aligned}
\end{equation}
where balancing parameters that weigh each term are empirically given as $\{ \omega_{\delta}, \omega_{1}, \omega_{0}, \omega_{C} \} = \{5, 1, 5, 3 \}$ in this study.

\subsection{Inference}
\label{subsec:inference}

\begin{figure}[!t]
\centerline{\includegraphics[scale = 0.32]{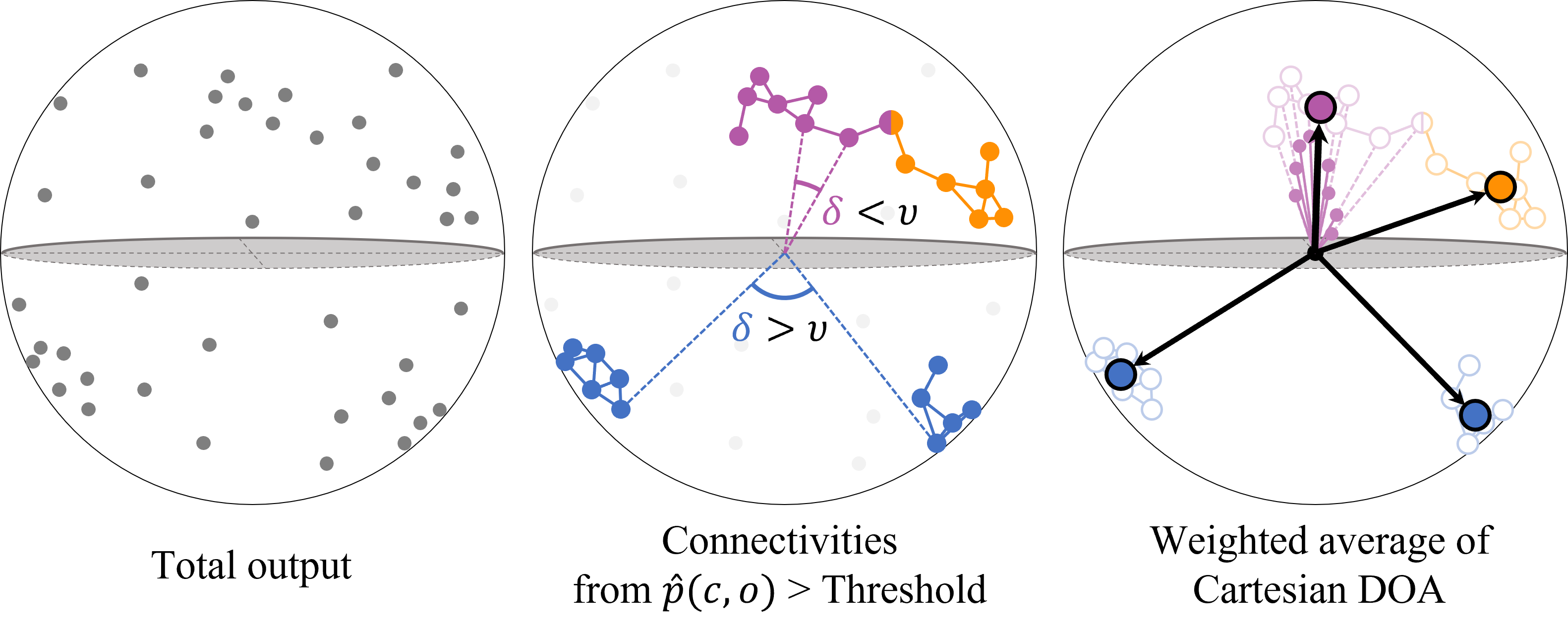}}
\caption{The procedure of the connectivity-based NMS at frame $t$ (from left to right). Each cluster comprises identical-class estimations where the same-color-solid edges connect the prediction group to be unified.}
\label{fig:NMS}
\end{figure}

Non-maximal suppression (NMS) can be adapted for the frame-wise SELD output using the angular distance measure.
Inspired by the ensemble technique suggested for SELD in \cite{Han_KU_task3_report}, connectivity-based unification is devised, clustering the predictions similar to the DBSCAN \cite{ester1996density} algorithm.
Considering the predictions of the conditional class probabilities $\hat{p}(c,o) = \hat{p}(c|o) \cdot \hat{p}(o)$ beyond the threshold of 0.5, we unify the predictions based on the estimated DOAs within the angular threshold $\upsilon$.
Subsequently, the weighted average of the Cartesian DOAs from the class-wise cluster determines the unified DOA estimation, where the weights are described as:
\begin{equation}
\small
\begin{aligned}
W = \underset{\forall i \in \text{cluster}}{\mathrm{softmax}}(\mathrm{exp}(\hat{p}_{i}(c,o)^{2} / \text{\space} 0.5))
\end{aligned}
\end{equation}
The weight vector $W \in \mathbb{R}^{|\text{cluster}|}$ comprises elements that are proportional to the confidence scores of each prediction.
An example of a connectivity-based NMS is shown in Fig. \ref{fig:NMS}.

\section{Experiments}
\label{sec:experiment}

\begin{table*}
\caption{Comparison of SELD performance with different representations on the series of DCASE Task 3 (SELD) development set from 2020 to 2022 competition. All formats used the same backbone network. $\mathrm{F}_{20^{\circ}}$ and $\mathrm{LR}_{\mathrm{CD}}$ are reported in percentages.}
\label{table:overall-performance}
\resizebox{1.0\textwidth}{!}{%
\begin{tabular}{lc | ccccc | ccccc | ccccc}
\toprule
\multirow{2}{*}{Format} & \multirow{2}{*}{\makecell{Unify \\ Threshold ($\upsilon$)}} &
    \multicolumn{5}{c}{DCASE 2022} \vline& \multicolumn{5}{c}{DCASE 2021} \vline& \multicolumn{5}{c}{DCASE 2020} \\
 & & $\mathrm{ER}_{20^{\circ}}^{\text{\space}\downarrow}$ & $\mathrm{F}_{20^{\circ}}^{\text{\space}\uparrow}$ & 
            $\mathrm{LE}_{\mathrm{CD}}^{\text{\space}\downarrow}$ & $\mathrm{LR}_{\mathrm{CD}}^{\text{\space}\uparrow}$ & $\varepsilon_{\mathrm{SELD}}^{\text{\space}\downarrow}$ &
     $\mathrm{ER}_{20^{\circ}}^{\text{\space}\downarrow}$ & $\mathrm{F}_{20^{\circ}}^{\text{\space}\uparrow}$ & 
            $\mathrm{LE}_{\mathrm{CD}}^{\text{\space}\downarrow}$ & $\mathrm{LR}_{\mathrm{CD}}^{\text{\space}\uparrow}$ & $\varepsilon_{\mathrm{SELD}}^{\text{\space}\downarrow}$ &
     $\mathrm{ER}_{20^{\circ}}^{\text{\space}\downarrow}$ & $\mathrm{F}_{20^{\circ}}^{\text{\space}\uparrow}$ & 
            $\mathrm{LE}_{\mathrm{CD}}^{\text{\space}\downarrow}$ & $\mathrm{LR}_{\mathrm{CD}}^{\text{\space}\uparrow}$ & $\varepsilon_{\mathrm{SELD}}^{\text{\space}\downarrow}$ \\
\midrule
SED-DOA \cite{adavanne2018sound} 
        & -            & 0.9619 & 11.40 & 31.01$^\circ$ & 42.18 & 0.6496 
                       & 0.6150 & 36.60 & 21.39$^\circ$ & 56.50 & 0.4507 
                       & {\bf 0.4768} & 59.87 & 9.50$^\circ$  & 65.17 & 0.3198 \\
                       
ACCDOA \cite{shimada2021accdoa}
        & -            & {\bf 0.7758} & 16.98 & 30.00$^\circ$ & 52.44 & 0.5621 
                       & 0.5872 & 40.31 & 20.65$^\circ$ & 63.36 & 0.4163 
                       & 0.5278 & 50.73 & 16.61$^\circ$ & 73.74 & 0.3439 \\
                       
ADPIT \cite{shimada2022multi}
        & $15^{\circ}$ & 0.8035 & 14.93 & 30.96$^\circ$ & 51.87 & 0.5769 
                       & 0.5990 & 39.50 & 19.63$^\circ$ & 64.49 & 0.4170 
                       & 0.5193 & 52.59 & 16.36$^\circ$ & {\bf 75.17} & 0.3332 \\
                       
        & $30^{\circ}$ & 0.7999 & 14.94 & 30.94$^\circ$ & 51.84 & 0.5760 
                       & 0.5894 & 39.70 & 19.61$^\circ$ & 64.39 & 0.4144 
                       & 0.5161 & 52.76 & 16.36$^\circ$ & 75.16 & 0.3320 \\
                       
        & $45^{\circ}$ & 0.7993 & 14.96 & 30.93$^\circ$ & 51.81 & 0.5759 
                       & 0.5887 & 39.70 & 19.60$^\circ$ & 64.35 & 0.4143 
                       & 0.5158 & 52.77 & 16.36$^\circ$ & 75.15 & 0.3319 \\   
\midrule
AD-YOLO    & $15^{\circ}$ & 0.8123 & 27.36 & {\bf 24.59$^\circ$} & {\bf 53.17} & 0.5359 
                          & 0.5897 & 52.42 & {\bf 13.45$^\circ$} & {\bf 65.25} & 0.3719 
                          & 0.5297 & 59.98 & {\bf 8.48$^\circ$}  & 69.82 & 0.3197 \\
                       
(proposed) & $30^{\circ}$ & 0.7870 & 27.54 & 24.63$^\circ$ & 52.71 & 0.5303 
                          & 0.5359 & 53.98 & {\bf 13.45$^\circ$} & 64.84 & 0.3556 
                          & 0.4927 & 60.94 & 8.56$^\circ$  & 69.81 & 0.3082 \\
                       
           & $45^{\circ}$ & 0.7802 & {\bf 27.59} & 24.69$^\circ$ & 52.62 & {\bf 0.5288} 
                          & {\bf 0.5185} & {\bf 54.35} & 13.54$^\circ$ & 64.70 & {\bf 0.3508} 
                          & 0.4818 & {\bf 61.27} & 8.60$^\circ$ & 69.75 & {\bf 0.3048} \\
\bottomrule
\end{tabular}}
\end{table*}

\subsection{Experimental setups}
\label{subsec:experimental-setups}

We adopted the development sets of DCASE Task 3 from 2020 to 2022 \cite{politis2020dataset, politis2021dataset, politis2022starss22} to compare the proposed method with other SELD approaches \cite{adavanne2018sound, shimada2021accdoa, shimada2022multi}.
Each includes 14, 12, and 13 sound event classes respectively, which are loosely shared.
The 2020 and 2021 data comprise one-minute sound scene emulation, where 400 training, 100 validation, and 100 test recordings are explicitly provided. 
Whereas polyphony $\leq 2$ was considered in 2020, the 2021 dataset contains polyphony $\leq 4$ recordings, including non-target interference.

The 2022 dataset comprises manually annotated real-world sound scenes between 30 s to 5 min duration. 
It contains 67 training and 54 test clips.
Real recordings are fairly common to multiple overlaps, where the maximum is five.
By considering the challenge objective in 2022, we exploited 1,200 simulated scenes\footnote{DOI: 10.5281/zenodo.6406873} as a training split, which was used for the baseline training of the year.
Correspondingly, the training and test splits of the 2022 development dataset were considered for validation and testing, respectively, in this setup.
The baseline data were emulated in the same manner as in 2020, where polyphony $\leq 2$ was allowed.

A first-order ambisonic format was used in the experiment, and every sound scene recording was sampled at 24 kHz.
For training, we segmented audio recordings into 20 s duration with 1 s shifting. 
Mel-spectrogram and intensity vectors were used as the network inputs, where the acoustic features were extracted using 50 ms Hanning window, 25 ms hop length, and 64 mel-scale bins.
Augmentations of spectrogram masking \cite{park19e_interspeech} and 16-pattern rotation \cite{mazzon2019first} were applied to train the same SE-ResNet\cite{hu2018squeeze}--BiGRU\cite{cho2014learning} architecture in \cite{Han_KU_task3_report}.
Adam optimizer of learning rate $1e^{-3}$ was adopted, and 500 iterations with 16 batch size learning were set in 200 training epochs.
For the AD-YOLO, we used a grid size $(45^{\circ},45^{\circ})$ with $50\%$ overlap extension and $\mathcal{T}=\{45^{\circ}, 25^{\circ}, 10^{\circ}\}$.
$\mathbf{K}$ from AD-YOLO and the number of tracks in ADPIT were set to $3$, the maximum polyphony of the training data.

\subsection{Evaluation metrics}
\label{subsec:evaluation-metrics}

We adopted four metrics for the evaluation \cite{mesaros2019joint}.
Considering the predictions as true positives when the angular distance to the reference was less than $20^{\circ}$, the location-sensitive error rate ($\mathrm{ER}_{20^{\circ}}$) and F-score ($\mathrm{F}_{20^{\circ}}$) were measured.
The localization recall ($\mathrm{LR}_{\mathrm{CD}}$) measured the class-wise true-positive ratio.
The class-sensitive localization error ($\mathrm{LE}_{\mathrm{CD}}$) averaged the angular distance between the references and the class predictions in degrees.
In addition, the SELD error $\varepsilon_{\mathrm{SELD}}=\mathrm{AVG}(\mathrm{ER}_{20^{\circ}}, 1-\mathrm{F}_{20^{\circ}}, 1-\mathrm{LR}_{\mathrm{CD}}, \mathrm{LE}_{\mathrm{CD}}/180^{\circ})$ averaged the errors in the aforementioned measurements.

\subsection{Experimental results}
\label{subsec:experimental-results}

Table \ref{table:overall-performance} presents the performances of several approaches to solving SELD problems.
In processing multiple predictions to address class-homogeneous overlapping, ADPIT and AD-YOLO were evaluated using multiple $\upsilon$ thresholds.
The variance in scores between datasets indicates that the different task setups significantly affected the system.
Nevertheless, AD-YOLO outperformed the other approaches in all setups, particularly in terms of $\mathrm{F}_{20^{\circ}}$ and $\mathrm{LE}_{\mathrm{CD}}$.

Table \ref{table:overlapping-performance} summarizes the performances evaluated exclusively for the same-class overlapping, and $\triangle\varepsilon_{\mathrm{SELD}}$ denotes the change from which in Table \ref{table:overall-performance}.
In the 2022 and 2020 setups, where only polyphony $\leq 2$ was offered in training, all formats experienced performance degradation.
AD-YOLO proved its robustness in class-homogeneous polyphony in terms of the minimum degradation of the performance.
Simultaneously, AD-YOLO outperformed the others in the absolute performance itself. 
In the 2021 setup, as more polyphonic ($\leq 4$) instances were trained and validated, the degradations were mitigated.
While the others still suffered from the degradation in this setup, AD-YOLO achieved the maintenance and demonstrated the best performance again.
\begin{table}[!t]
\centering
\caption{SELD performance evaluation on overlapping sound events of the same class.}
\label{table:overlapping-performance}
\resizebox{0.473\textwidth}{!}{%
\begin{tabular}{lc|c|c|c}
\toprule
\multirow{2}{*}{Format} & \multirow{2}{*}{\makecell{Unify \\ Threshold ($\upsilon$)}} &
        \multicolumn{3}{c}{$\varepsilon_{\mathrm{SELD}}{\downarrow}$\footnotesize{ ($\triangle\varepsilon_{\mathrm{SELD}}$)}} \\
 & & DCASE 2022 & DCASE 2021 & DCASE 2020 \\
\midrule
SED-DOA \cite{adavanne2018sound}
        & -            & 0.7730\footnotesize{ (\textcolor{red}{+0.12})} & 0.5259\footnotesize{ (\textcolor{red}{+0.08})} & 0.6713\footnotesize{ (\textcolor{red}{+0.35})} \\
ACCDOA  \cite{shimada2021accdoa}
        & -            & 0.7742\footnotesize{ (\textcolor{red}{+0.21})} & 0.4691\footnotesize{ (\textcolor{red}{+0.05})} & 0.6608\footnotesize{ (\textcolor{red}{+0.32})} \\
ADPIT   \cite{shimada2022multi}
        & $15^{\circ}$ & 0.7315\footnotesize{ (\textcolor{red}{+0.15})} & 0.4632\footnotesize{ (\textcolor{red}{+0.05})} & 0.6985\footnotesize{ (\textcolor{red}{+0.37})} \\
        & $30^{\circ}$ & 0.7316\footnotesize{ (\textcolor{red}{+0.16})} & 0.4593\footnotesize{ (\textcolor{red}{+0.04})} & 0.6999\footnotesize{ (\textcolor{red}{+0.37})} \\
        & $45^{\circ}$ & 0.7313\footnotesize{ (\textcolor{red}{+0.16})} & 0.4590\footnotesize{ (\textcolor{red}{+0.04})} & 0.7002\footnotesize{ (\textcolor{red}{+0.37})} \\
\midrule
AD-YOLO    & $15^{\circ}$ & {\bf 0.6011}\footnotesize{ (\textcolor{red}{\bf +0.07})} & 0.3654\footnotesize{ (\textcolor{blue}{\bf -0.01})} & 0.5623\footnotesize{ (\textcolor{red}{\bf +0.24})} \\
(proposed) & $30^{\circ}$ & 0.6093\footnotesize{ (\textcolor{red}{+0.08})} & {\bf 0.3613}\footnotesize{ (\textcolor{red}{+0.01})} & {\bf 0.5557}\footnotesize{ (\textcolor{red}{+0.25})} \\
           & $45^{\circ}$ & 0.6142\footnotesize{ (\textcolor{red}{+0.09})} & 0.3623\footnotesize{ (\textcolor{red}{+0.01})} & 0.5571\footnotesize{ (\textcolor{red}{+0.25})} \\
\bottomrule
\end{tabular}}
\end{table}

\subsection{The influences of SELD adaptations in AD-YOLO}
\label{subsec:ablation}
Table \ref{table:ablation} presents the influences of the main SELD adaptations of AD-YOLO.
The first was to regress the angular loss to the mean squared error (MSE) of the relative coordinates as the YOLO.
The other removed the sound existence losses $l_{[0,1];\tau}$, such that the class scores managed the presence of each event.
The results show that both adaptations significantly affected the AD-YOLO performance.
Furthermore, the result from 2021 suggests that learning the condition when the target class exists has a major role in AD-YOLO under interference.
\begin{table}[!t]
\centering
\caption{Comparison results depending on each AD-YOLO SELD adaptation. The predictions were unified in $\upsilon=15^{\circ}$.}
\label{table:ablation}
\resizebox{0.473\textwidth}{!}{%
\begin{tabular}{l|c|c|c}
\toprule
\multirow{2}{*}{w/o. Adaptation} &
        \multicolumn{3}{c}{$\varepsilon_{\mathrm{SELD}}{\downarrow}$} \\
  & DCASE 2022 & DCASE 2021 & DCASE 2020 \\
\midrule
$l_{\delta;\tau} \text{\space\space\space\space} \rightarrow \mathrm{MSE}_{(\lambda, \phi; \tau)}$ 
    & 0.5743 & 0.4288 & 0.3559 \\
$l_{[1,0];\tau}  \rightarrow$ None   
    & 0.5824 & 0.7348 & 0.3664 \\
\midrule
AD-YOLO 
    & \bf0.5359 & \bf0.3719 & \bf0.3197 \\
\bottomrule
\end{tabular}}
\end{table}

\section{Conclusion}
\label{sec:conclusion}

We proposed an angular-distance-based YOLO (AD-YOLO) approach to perform sound event localization and detection (SELD) on a spherical surface.
AD-YOLO assigns multi-layered responsibilities, which are based on the angular distance from the target events, to predictions according to each estimated direction of arrival.
Avoiding the primal format of the event-oriented track output, AD-YOLO addresses the SELD problem in an unknown polyphony environment. 
In a series of evaluations on the data given in DCASE 2020 to 2022 Task 3, the model trained using the AD-YOLO approach exhibits outstanding performance and also proves its robustness in a class-homogeneous polyphony environment.

\vfill\pagebreak
\bibliographystyle{IEEEbib-abrv}
\bibliography{paper}

\end{document}